\documentclass{emulateapj}

\begin{document}
\title{Early optical-IR emission from GRB 041219a:
neutron-rich internal shocks and a mildly magnetized
external reverse shock} 

\author{Y. Z. Fan$^{1,2,3}$, Bing Zhang$^{3}$ and
 D. M. Wei$^{1,2}$}
\affil{$^1$ Purple Mountain Observatory, Chinese Academy of
Science, Nanjing 210008, China.\\
$^2$ National Astronomical Observatories, Chinese Academy of
Sciences, Beijing, 100012, China.\\
$^3$ Dept. of Physics, University of Nevada, Las Vegas, NV
89154, USA.\\ } 

\begin{abstract}
Very early optical and near infrared (IR) emission was discovered
accompanying the long gamma-ray burst (GRB) 041219a. We show that the
optical/IR flash tracking the gamma-ray lightcurve during the prompt
emission could be understood as emission from neutron-rich internal
shocks, as has been suggested by Fan \& Wei. The early $K_{\rm
s}$-band afterglow lightcurve after the prompt phase could be
well-modeled as the superposition of a reverse shock and a forward
shock component. The
radio data also support the reverse shock interpretation. 
\end{abstract}


\section{Introduction}
GRB 041219a was detected both by the IBIS (Imager on Board the
INTEGRAL Satellite) detector of the INTEGRAL satellite (Gotz et
al. 2004) and by the Swift Burst Alert Telescope (BAT) (Barthelmy et
al. 2004). This burst distinguishes itself from other bursts in
several aspects. (1) It is very bright. The 15-350 keV fluence measured
by the Swift BAT was approximately $1.55\times 10^{-4}~{\rm
ergs/cm^2}$ (Barthelmy et al. 2004), placing it in the top few
percent among the 1637 GRB events listed in the comprehensive fourth
{\em Burst and Transient Source Experiment} (BATSE) catalogue
(Paciesas et 
al. 1999). (2) The duration of prompt $\gamma-$ray emission ($T_{90}$)
is approximately 520 seconds, making it one of the longest bursts ever
detected. (3) The prompt optical and infrared (IR) emission was
detected to accompany the prompt $\gamma-$ray emission (Vestrand et
al. 2005; Blake et al. 2005). This is the second case since GRB 990123
(e.g., Akerlof et al. 1999).  (4) The location of this burst is near
the Galactic plane and in a direction with high optical extinction
(Galactic coordinates ${\rm l}=120^0$, $b=+0.1^0$), so that the R-band
extinction is very large ($\sim 4.9$ magnitudes or even larger). For
this reason, no late optical afterglow has been detected. Fortunately,
in the IR band, the afterglow (including that in the very early phase)
has been well detected (Blake et al. 2005). The redshift is
unknown. In this Letter, we assume $z=1$.

The very early long-wavelength observation is very important to
diagnose the outflow composition (Fan et al. 2004; Zhang \& Kobayashi
2005; Fan et al. 2005)\footnote{In principle, the ejecta of
GRBs may be Poynting flux dominated (see Lyutikov \& Blandford 2003
and the references therein) or neutron-rich (e.g., Derishev et
al. 1999; Beloborodov 2003; Pruet et al. 2003) or both (e.g., Vlahakis
et al. 2003).},
since the late
afterglow taking place hours after the burst trigger is powered by the
forward shock (FS), so that essentially all the initial information of
the ejecta is lost. 
In this Letter, we apply our previous analyses to 
discuss the very early long-wavelength observation of
GRB 041219a. 

\section{The prompt optical and near-IR flash}\label{Prompt}  
There has been some interest in discussing/searching for prompt
long-wavelength radiation accompanying prompt $\gamma-$rays 
emission even in the pre-afterglow era (e.g., Katz 1994; Schaefer et 
al. 1994; Wei \& Cheng 1997).  In the afterglow era, more 
theoretical attention was paid on the topic. In the standard 
internal shocks model, accompanying the prompt $\gamma-$ray emission, 
long wavelength flashes are expected (M\'{e}sz\'{a}ros \&
Ress 1997, 1999; Fan \& Wei 2004a).
If there are a large amount of neutrons contained in the GRB
outflow, the decayed neutron shells would provide more collisions at
a larger distance from the central engine. If a burst is long
enough, the neutron-rich internal shocks would give rise to detectable
long wave-length flashes during the prompt $\gamma-$ray emission phase
(Fan \& Wei 2004b).

In the standard internal shock model, the
synchrotron-self-absorption frequency could be estimated as (e.g., Li
\& Song 2004) 
$\nu_{\rm a}\sim 7\times 10^{16}{\rm Hz}~L_{\rm syn,52}^{2/7}\Gamma_{\rm
m,2.5}^{3/7}R_{\rm int,13}^{-4/7}{B'}_{\rm 4}^{1/7}({2\over 1+z})$,
where $L_{\rm syn}$ is the synchrotron radiation luminosity, $B'$ is
the comoving-frame magnetic field strength in the internal shock
phase, $R_{\rm int}\sim 2 \Gamma_{\rm m}^2c\delta t/(1+z)$ is the
typical internal shock radius, $\Gamma_{\rm m}$ is the bulk LF of the
merged two shells, $\delta t$ is the observed variable timescale of
the prompt $\gamma-$ray lightcurve, and the convention $Q_{\rm
x}=Q/10^{\rm x}$ has been adopted in cgs units here and throughout the
text. For typical GRB parameters, one can see that $\nu_{\rm a}$ is
usually well above the optical band. 
This tends to suppress the optical flux.
Also the $F_\nu$ spectrum is expected to have a power law index 5/2,
inconsistent with the prompt IR data of GRB
041219a (see Figure. 2 of Blake et al. 2005). Although with proper
adjustment of parameters, the proton-dominated internal shocks models
may be able to match the observation, in this Letter we focus an
alternative interpretation, i.e. the neutron-rich internal shock model
(Fan \& Wei 2004b). 

There are good reasons to assume that a large amount of neutrons
(comparable to the amount of protons) exist in the GRB ejecta (e.g.,
Derishev et al. 1999; Beloborodov et al. 2003; Pruet et al. 2003). In
the neutron-rich internal shock model, the LFs of the proton shells
are variable, so are the LFs of accompanying neutron shells.  For the
slow neutrons (with $\Gamma_{\rm n,s} = 50$) coupled with the slow
proton shells, they do not interact with other materials before
decaying (the typical $\beta-$decay radius is $\sim 900~{\rm s}~ c
\Gamma_{\rm n,s} = 1.3\times 10^{15}$ cm, where $c$ is the speed of
light).  However, if the GRB lasts long enough, the slow neutron
shells ejected at earlier times would be swept successively by the
faster proton shells ejected at later times. This happens in the
distance range of $\sim 10^{13}-{\rm several} \times 10^{15}$ cm, in
which the slow neutron shells decay continuously. The proton shells
interact with the $\beta-$decay products of the slow neutron shells
and power detectable long wavelength prompt emission, as shown in Fan
\& Wei (2004b).
            
A detailed treatment of the process has been presented in \S{2} of Fan
\& Wei (2004b). A novel effect taken into account here is
the inverse Compton (IC) cooling of the electrons because of the
space-time overlapping between the proton shell$-$neutron shell
interaction region and the prompt MeV $\gamma-$ray photon flow
(e.g. Beloborodov 2005; Fan et al. 2005).  The inverse Compton
parameter is calculated as $Y= P_{_{\rm IC}}/P_{\rm syn}$, where
$P_{_{\rm IC}}$ is the energy loss rate of one electron via IC scattering
with the prompt $\gamma-$ray emission, $P_{\rm syn}=(4/3)\sigma_{\rm
T}\gamma_{\rm e}^2\beta_{\rm e}^2U_{\rm B}c$ is the energy loss rate
of one electron via synchrotron radiation, $\gamma_{\rm e}$ is the
random Lorentz factor of the emitting electron, and $U_{\rm B}$ is the
magnetic energy density generated during the interaction.  Since the
Klein-Nishina correction is important for the IC process discussed
here, we strictly use Eqs. (2.47-2.51) and Eq. (2.56) in 
Blumenthal \& Gould (1970) to calculate $P_{_{\rm IC}}$. The prompt
$\gamma-$rays photon number distribution is taken as
$n_{\epsilon_\gamma}=n_{\epsilon_\gamma^{\rm
b}}(\epsilon_\gamma/\epsilon_\gamma^{\rm b})^{-3/2}$ for
$\epsilon_\gamma^{\rm b}/50<\epsilon_\gamma<\epsilon_\gamma^{\rm b}$
and $n_{\epsilon_\gamma}=n_{\epsilon_\gamma^{\rm
b}}(\epsilon_\gamma/\epsilon_\gamma^{\rm b})^{-2.25}$ for
$\epsilon_\gamma^{\rm b}<\epsilon_\gamma<50\epsilon_\gamma^{\rm b}$,
where $\epsilon_\gamma^{\rm b}\sim 250$ keV is the peak energy of the
observed spectrum\footnote{Given a same $\epsilon_\gamma^{\rm b}$, the
calculated results are not very sensitive to the photon spectral indices
assumed.}, $n_{\epsilon_\gamma^{\rm b}}$ is constrained by the
condition $\int n_{\epsilon_\gamma}\epsilon_\gamma d \epsilon_\gamma
=U_\gamma$, where $U_\gamma\approx L_\gamma/4\pi R^2 \gamma^2 c$ is
the initial $\gamma-$ray photon energy density, $\gamma$ is the
evolving bulk LF of the proton shell during the sweeping
process\footnote{Notice that we have adopted the same symbols as in
Fan \& Wei (2004b) to keep consistency between the two papers.}, and
$L_\gamma$ is the prompt $\gamma-$ray luminosity.

\begin{figure}
\epsscale{1.0}
\plotone{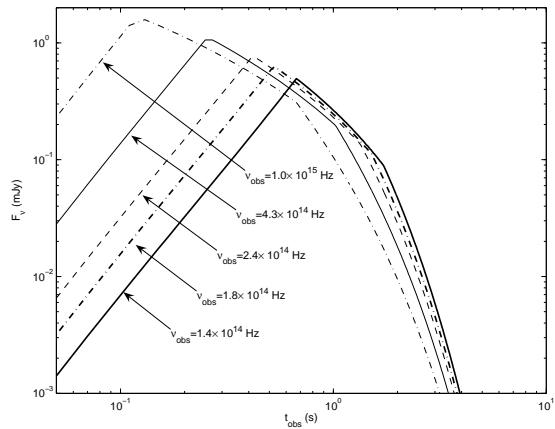}
\caption{Long wavelength (the observer frequency has been marked in
the figure) emission powered by a proton shell interacting with the
decay products of a series of slow neutron shells as a function
of time (Similar to Fig. 2 of Fan \& Wei (2004b)). Following
parameters are taken:  $M_{\rm p}=3.7\times 10^{26}$g, $M_{\rm
n}^0=10^{26}$g, $\Gamma_{\rm m}=600$, $\Gamma_{\rm n,s}=50$, $\delta
t=10^{-2}(1+z)$s, $D_{\rm L}=2.2\times 10^{28}$cm, $\varepsilon_{\rm
e}=0.3$, $\varepsilon_{\rm B}=0.1$,  $L_\gamma=10^{52}{\rm
ergs~s^{-1}}$ and $\epsilon_\gamma^{\rm b}=250$keV. The real
lightcurve should include the contributions from many proton shells.}
\label{FW04b} 
\end{figure}

We calculated the synchrotron lightcurves for a proton shell
interacting with the decay products of a series of slow neutron
shells. The result is shown in 
Fig. \ref{FW04b}. Following parameters have been taken: $M_{\rm
p}=3.7\times 10^{26}$g is the rest mass of one proton shell; $M_{\rm
n}^0=10^{26}$g is the initial rest mass of one neutron shell;
$\Gamma_{\rm m}=600$; $\Gamma_{\rm n,s}=50$ (With these parameters, we
get $\gamma \simeq 300$ at the end of the interaction. This will be
regarded as the intial Lorentz factor in the afterglow phase, and we
will rewrite it as $\eta$ in \S{\ref{Afterglow}}. This matches the one
found in our modeling the early $K_{\rm s}$ band afterglow. See
\S{\ref{Afterglow}}); $\delta t=10^{-2}(1+z)$s; $D_{\rm L}=2.2\times
10^{28}$cm is the luminosity distance; $L_\gamma=10^{52}{\rm
ergs~s^{-1}}$; and $\varepsilon_{\rm e}=0.3$ and $\varepsilon_{\rm B}=0.1$
are the fraction of shock energy given to the shocked electrons and
magnetic field, respectively. The lightcurves reach the peak as the
observer frequency $\nu_{\rm obs}$ crosses the
synchrotron-self-absorption frequency $\nu_{\rm a}$, and another break
during the decaying phase marks the epoch as  
$\nu_{\rm obs}$ crosses the typical synchrotron radiation
frequency. The $\nu_{\rm a}$ is much smaller than that
in the standard internal shock model
because of the following reasons. Firstly, we have $R\gg R_{\rm int}$.
Secondly, the forward shock upstream proton number density is much
smaller than that in the standard internal shocks (see eq.[1] of Fan
\& Wei (2004b) for detail), which results in a smaller $B'$ and weaker
synchrotron emission. Finally, with the IC cooling effect taken into
account, and since synchrotron emission is weaker, 
the $Y$ parameter is large, i.e. about tens
to hundreds with the typical parameters. The synchrotron radiation
luminosity is further lowered by a factor $1/Y$.

The detected flux should be the integrated emission powered by a
series of proton shells interacting with the decay trail of slow
neutron shells. For $\nu_{\rm obs}=(1.4,~1.8,~2.2,~4.3,~10.0)\times
10^{14}$ Hz, the predicted flux are
$(18.8~,21.2~,~23.3,~26.3,~27.2)$ mJy, respectively. Such strong
emission is detectable with the current telescopes, and is consistent
with the optical observation of GRB 041219a when the extinction
correction is taken into account (Vestrand et al. 2005). The averaged
spectrum is flat, which is roughly consistent with the earliest IR
band observation of GRB 041219a (Blake et al. 2005). As mentioned in
Fan \& Wei (2004b), the predicted long-wavelength emission is expected
to be correlated with the prompt $\gamma-$ray emission but has a $\sim
10(1+z)\Gamma_{\rm n,s,1.7}^{-1}$ s lag. This is also consistent with
the tracking behavior of the 
prompt optical flashes (Vestrand et al. 2005). 
Notice that it is not our intention to fit the prompt optical and near
IR emission lightcurves closely (which requires additional complicated
assumptions about the central engine behavior). The main purpose of the
current discussion is to indicate that the neutron-rich internal shock
model suffers less constraints than the proton-dominated internal shock
model, and can better account for the observed prompt optical/IR
spectrum. 

\section{The very early near-IR afterglow}\label{Afterglow} 
After the internal shock phase, as the fireball is decelerated by the
circumburst medium, usually a pair of shocks develop (M\'esz\'aros \&
Rees 1997; Sari \& Piran 1999). The early optical afterglow lightcurve
is usually composed of the contributions from both the forward and the
reverse shocks. Zhang et al. (2003) pointed out
that depending on parameters, there 
are two types of early optical/IR lightcurves for a fireball
interacting with a constant density medium (ISM), i.e. Type I
(rebrightening) in which distinct reverse shock and the forward
shock peaks are detectable, and Type II (flattening) in which the
forward shock peak is buried beneath the reverse shock peak. The
previous two strong cases of reverse shock emission (GRB 990123,
Akerlof et al. 1999; and GRB 021211, Fox et al. 2003, Li et al. 2003)
all belong to Type II. Visual inspection of the early IR
lightcurve of GRB 021219a (Blake et al. 2005, see also Fig.{\ref{NIR})
indicates that it is a clear Type I case.
Below we will model the lightcurve in detail and show that the data
are indeed consistent with such an explanation.

Following the
standard afterglow model for a fireball interacting with a constant
density medium (e.g., Piran 1999), we write down the cooling
frequency $\nu_{\rm c}^{\rm f}$, the typical synchrotron frequency
$\nu_{\rm m}^{\rm f}$ and the maximum spectral flux $F_{\rm
\nu,max}^{\rm f}$ of the FS emission, i.e., $
\nu_{\rm c}^{\rm f}=1.4\times 10^{14}{\rm
Hz}~E_{\rm iso,54}^{-1/2}\varepsilon_{\rm B,-2}^{-3/2}n_0^{-1}{t}_{\rm
d}^{-1/2}[2/(1+z)]$, 
$\nu_{\rm m}^{\rm f}=1.8\times 10^{13}{\rm
Hz}~E_{\rm iso, 54}^{1/2}\varepsilon_{\rm B,-2}^{1/2}\varepsilon_{\rm
e,-0.5}^2{t}_{\rm d}^{-{3/2}}C_{\rm p}^2[2/(1+z)]$, 
$F_{\rm \nu,max}^{\rm f}=83 {\rm mJy}~E_{\rm iso,54}\varepsilon_{\rm
B,-2}^{1/2}n_0^{1/2}D_{\rm L,28.34}^{-2}[(1+z)/2]$,
where $C_{\rm p}=7(p-2)/[2(p-1)]$, $E_{\rm iso}$ is the isotropic
energy of the outflow, $\varepsilon_{\rm e}$ and $\varepsilon_{\rm B}$
are the fractions of the shock energy given to electrons and to
magnetic fields in the forward shock, respectively, $n$ is the number
density of the external medium, $p\sim2.4$ is the power-law
distribution index of the shocked electrons. Hereafter $t=t_{\rm
obs}/(1+z)$ denotes the observer's time corrected for the cosmological
time dilation effect, and $t_{\rm d}$ is in unit of day.  The
superscript ``f'' (``r'') represent the FS (RS) emission,
respectively. We assume that $\varepsilon_{\rm e}$ and the electron
spectral index $p$ are essentially the same for both the FS and
RS\footnote{If they are different, additional corrections are needed,
see Zhang et al. (2003) and Zhang \& Kobayashi (2005) for details.},
but we will allow different $\varepsilon_{\rm B}$ values for both
regions. One reason for this assumption is that the magnetic
field generated in the internal shock phase may have not been
dissipated effectively in a short time, and would play a dominant role
in the reverse shock region (Fan et al. 2004 and the references listed
therein). In this 
Letter, we will denote $\varepsilon_{\rm B}^{\rm f}=\varepsilon_{\rm B}$
and $\varepsilon_{\rm B}^{\rm r}={\cal R_{\rm B}^{\rm
2}}\varepsilon_{\rm B}$, where ${\cal R}_{\rm B}$ is the ratio of the
magnetic field in the RS emission region to that in the FS emission
region (Zhang et al. 2003). Previous analyses indicate that at least
for some bursts (e.g. GRB990123 and GRB021211) the RS emission region
is more magnetized than the FS region 
(e.g. Fan et al. 2002; Zhang et al. 2003; Kumar \& Panaitescu 2003).

As shown in Fig.\ref{NIR} (data taken from Blake et al. (2005), only
the richest $K_{\rm s}$-band data are plotted),
the time when the RS crosses the
ejecta ((1+z)$t_\times$) is about 30 minutes after the trigger, which
is much longer than $T_{\rm 90}\sim 520$s.  So, the RS is
non-relativistic (e.g., Sari \& Piran 1995; Kobayashi 2000; Kobayashi
\& Zhang 2003). Further evidence for a non-relativistic RS is the
rapid increase of the early afterglow lightcurve (e.g., Kobayashi
2000; Kobayashi \& Zhang 2003). 
$\gamma_{34,\times}\approx (\eta/\Gamma_\times+\Gamma_\times/\eta)/2$,
the LF of the decelerated outflow relative to the initial LF at
$t_\times$, can be estimated by solving 
$cdt/ d\Delta \approx
(1-\beta_{\Gamma_3})\{1-\eta/
[\Gamma_3 (4\gamma_{34}+3)]\}/(\beta_{\eta}-\beta_{\Gamma_3})$ (e.g., Sari \&
Piran 1995; Fan et al. 2004) numerically.
Here $\Gamma_3$ is the bulk
LF of the shocked ejecta, $\eta$ is the initial bulk LF of the
outflow, $\Gamma_\times$ is the LF of the 
decelerated ejecta at $t_\times$, $\beta_{\rm A}$ is the corresponding
velocity (in unit of $c$) of the LF $\Gamma_{\rm A}$.  
For GRB 041219a, one has $t_\times \sim 4 T_{90}/(1+z)$, so that
$\gamma_{34,\times}-1$ is much smaller than 1, i.e., the RS is
non-relativistic. In such a case,
$t_\times$ can be approximated as
\begin{equation}
t_\times \approx 64{\rm s}~E_{\rm iso,54}^{1/3}n_0^{-1/3}\eta_{2.5}^{-8/3}.
\end{equation}
The typical frequency of the RS emission is
\begin{equation}
\nu_{\rm m}^{\rm r}(t_{\rm \times})={\cal R}_{\rm
B}(\gamma_{34,\times}-1)^2 
\nu_{\rm m}^{\rm f}(t_{\rm \times})/(\Gamma_{\times}-1)^2,
\end{equation}
Following Zhang et al. (2003), we have
\begin{eqnarray}
\nu_{\rm c}^{\rm r}\approx {\cal R}_{\rm B}^{-3}\nu_{\rm c}^{\rm
f},~~~F_{\rm \nu, max}^{\rm r}(t_{\rm \times})\approx \eta {\cal
R}_{\rm B} F_{\rm \nu, max}^{\rm f}(t_{\rm \times}).
\end{eqnarray}
Generally, the $K_{\rm s}$ band flux satisfies $F_{\nu_{\rm K_{\rm
s}}}(t_\times)\approx F_{\rm \nu,max}^{\rm r}(t_\times)[\nu_{\rm
K_{\rm s}}/\nu_{\rm m}^{\rm r}({\rm t_\times})]^{\rm -(p-1)/2}$.

\begin{figure}
\epsscale{1.0}
\plotone{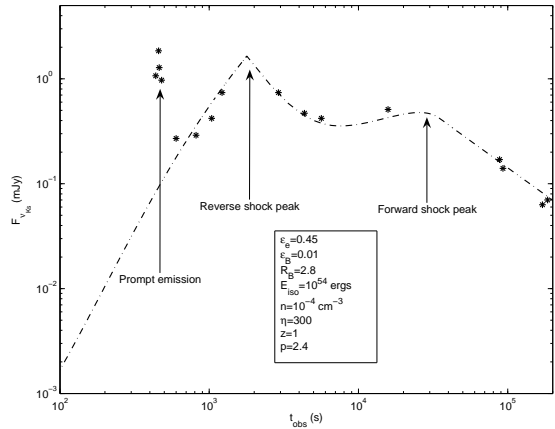}
\caption{Modeling the $K_{\rm s}$-band afterglow data of GRB
041219a. The data are taken from Blake et al. (2005). The earliest
data are prompt emission, which are not included in our fit. The
dash-dotted line is the theoretical lightcurve. Both the reverse
shock and the forward shock emission components are included.  The
best fit parameters are  marked in the text
box. }  
\label{NIR}
\end{figure}

Assuming $z=1$, $p=2.4$ and $E_{\rm iso,54}=1$ (consistent with the
gamma-ray fluence and our assumed redshift), $\varepsilon_{\rm e}$,
$n$ and $\eta$ can be constrained by the following two conditions. (1)
At $t_{\rm d}\sim 0.14$, $\nu_{\rm m}^{\rm f}\sim \nu_{\rm K_{\rm s}}$
and $F_{\rm \nu,max}^{\rm f}\sim 0.6$ mJy; (2) $t_\times$ is about
30/(1+z) minutes. We then get
\begin{equation}
\varepsilon_{\rm e}\sim 0.2\varepsilon_{\rm B,-2}^{-1/4},~~n\sim
5\times 10^{-5}\varepsilon_{\rm B,-2}^{-1},~~\eta\sim
380\varepsilon_{\rm B,-2}^{1/8}. 
\label{Constraint}
\end{equation}
It is interesting to see that since $\varepsilon_{\rm e}<1$, we have
$\varepsilon_{\rm B}>1.6\times 10^{-5}$, $n<3\times 10^{-2}~{\rm cm^{-3}}$
and $\eta>170$.  The parameter ${\cal R}_{\rm B}$ can be constrained
by noticing $F_{\nu_{\rm K_{\rm s}}}[(1+z)t_\times]\sim 2.4$ mJy and by
taking $\eta\sim 380$, which reads
\begin{equation}
{\cal R}_{\rm B}\sim 3[(\gamma_{34,\times}-1)/0.2]^{-14/17},
\end{equation}
which hints that the reverse shock region is mildly magnetized.

Following Fan et al. (2005), the forward-reverse shock emission has
been calculated numerically. Since the RS is non-relativistic, the
spreading of the ejecta (e.g. Piran 1999) has been taken into
account. The fits to the $K_{\rm s}$ band data are
presented in Fig. \ref{NIR}. It is found that the data can be
well modeled with the following parameters: $z=1$, $E_{\rm
iso,54}=1$, $\varepsilon_{\rm e}=0.45$, $\varepsilon_{\rm B}=0.01$,
${\cal R}_{\rm B}=2.8$, $\eta=300$, $n=10^{-4}~{\rm cm^{-3}}$. These
are consistent with the analytical estimates above.

There are three radio data points available (Soderberg \& Frail 2004;
van der  Horst et al. 2004a, 2004b). The 8.5GHz flux at 1.1 day is
$0.45$mJy, and the 4.9 GHz flux at 1.6 day and 2.6 day are 0.2mJy and
0.34mJy, respectively. By taking our best fitted parameters, the
corresponding FS fluxes are $\sim (0.04,~0.04,~0.05)$mJy,
respectively, 
too low to interpret the data. When we consider the RS contribution to
the radio band (e.g. Sari \& Piran 1999; Kobayashi \& Zhang 2003; Gou
et al. 2004), 
the over all corresponding fluxes become $\sim (0.60,~0.37,~0.18)$mJy,
respectively, roughly matching the data.
Therefore, the radio data also support the reverse-forward shock 
interpretation.

\section{Summary \& Discussion}

The prompt optical/IR observations of GRB 041219a (Vestrand et
al. 2005; Blake et al. 2005) offer a great opportunity to diagnose the
unknown GRB ejecta composition. We have shown that the prompt optical
emission tracking the gamma-ray emission profile may be consistent
with the picture that the ejecta is neutron rich, and that the optical
emission is powered by the proton shells interacting with the neutron
decay products at a distance farther away from the central engine than
the typical internal shock radius.

By modeling the $K_{\rm s}$ band early afterglow lightcurve, we
identify a reverse shock emission component, which is clearly
separated from the forward shock emission component.  Such a
rebrightening (Type I) lightcurve has been expected by Zhang et
al. (2003) to be more common if the RS is not strongly magnetized. Indeed,
detailed modeling indicates that ${\cal R}_{\rm B}$ in GRB 041219a is at most 
 mild, in contrast with GRB 990123 and GRB 021211 (e.g., Fan et
al. 2002; Zhang et al. 2003; Kumar \& Panaitescu 2003). This is also
consistent with the neutron-rich picture conjectured in interpreting
the prompt optical emission. The mild magnetization of the ejecta may
be due to magnetic field generation during the internal shock phase.
The radio data also support our reverse-forward shock interpretation.

It is interesting to note that for the three bursts with reverse shock
identification (GRB 990123, GRB 021211 and GRB 041219a), the inferred
number densities of interstellar medium are typically lower than the
standard value for ISM $n\sim 1~{\rm 
cm^{-3}}$ (see also Kumar 2004) --- For GRB 990123, $n\sim 10^{-3}~{\rm
cm^{-3}}$ (e.g., Wang 
et al. 2000; Panaitescu \& Kumar 2002; Nakar \& Piran 2004); For GRB
021211, $n\sim 10^{-3}-10^{-2}~{\rm cm^{-3}}$ (e.g., Wei 2003; Kumar \&
Panaitescu 2003); For the current GRB 041219a, $n \sim 10^{-4}~{\rm
cm^{-3}}$. The reason is unclear, but certain selection effects may
play a role.

\acknowledgments We thank L. J. Gou and X. F. Wu for helpful discussions, 
and the referee for valuable comments.
This work is supported by NASA NNG04GD51G and a NASA Swift GI (Cycle
1) program (for B.Z.), the National Natural Science Foundation (grants
10073022, 10225314 and 10233010) of China, and the National 973 
Project on Fundamental Researches of China (NKBRSF G19990754) (for
D.M.W.).

\end{document}